\documentclass[preprint,superscriptaddress]{revtex4-1}

\usepackage{soul}
\usepackage{color}
\usepackage{amsmath}
\usepackage{graphicx}
\usepackage{epstopdf}
\usepackage{mathtools}
\usepackage{textgreek}

\begin{document}
	\title{Sensing with exceptional surfaces: combining sensitivity with robustness}
	\author{Q. Zhong}
		\affiliation{Department of Physics and Henes Center for Quantum Phenomena, Michigan Technological University, Houghton, Michigan, 49931, USA}
		
	\author{J. Ren}
	\affiliation{College of Optics $\&$ Photonics-CREOL, University of Central Florida, Orlando, Florida, 32816, USA}

	\author{M. Khajavikhan}
	\affiliation{College of Optics $\&$ Photonics-CREOL, University of Central Florida, Orlando, Florida, 32816, USA}
	
	\author{D.N. Christodoulides}
	\affiliation{College of Optics $\&$ Photonics-CREOL, University of Central Florida, Orlando, Florida, 32816, USA}
	
	\author{\c{S}.K. {\"O}zdemir }
	\affiliation{Department of Engineering Science and Mechanics, The Pennsylvania State University, University Park, Pennsylvania 16802, USA}
	
	\author{R. El-Ganainy}
	\email[]{ganainy@mtu.edu}
	\affiliation{Department of Physics and Henes Center for Quantum Phenomena, Michigan Technological University, Houghton, Michigan, 49931, USA}

\begin{abstract}
	Exceptional points (EPs) are singularities that arise in non-Hermitian physics. Current research efforts focus only on systems supporting isolated EPs characterized by increased sensitivity to external perturbations, which makes them potential candidates for building next generation optical sensors. On the downside,  this feature is also the Achilles heel of these devices: they are very sensitive to fabrication errors and experimental uncertainties. To overcome this problem, we introduce a new design concept for implementing photonic EPs that combine the robustness required for practical use together with their hallmark sensitivity. Particularly, our proposed structure exhibits a hypersurface of Jordan EPs (JEPs) embedded in a larger space, and having the following peculiar features: (1) A large class of undesired perturbations shift the operating point along the exceptional surface (ES), thus leaving the system at another EP which explains the robustness; (2) Perturbations due to back reflection/scattering force the operating point out of the ES, leading to enhanced sensitivity. Importantly, our proposed geometry is relatively easy to implement using standard photonics components and the design concept can be extended to other physical platforms such as microwave or acoustics.   
\end{abstract}
\maketitle

\section*{Introduction}
Exceptional points (EP) are peculiar singularities that arise in non-Hermitian Hamiltonians when two or more eigenstates coalesce \cite{Heiss1990JPA,Magunov1999JPB,Heiss2004JPA,Heiss2012JPA,Rotter2003PRE,Muller2008JPA}. The resultant reduction in the eigenstate space dimensionality renders these points very sensitive to any external perturbations. Current research works in non-Hermitian and parity-time (PT) symmetric physics \cite{El-Ganainy2018NP,Feng2017NP}  have so far focused on systems supporting isolated EPs in a reduced parameter space. This strategy has allowed researchers to investigate certain important aspects of non-Hermitian systems and gain insight into their behavior \cite{El-Ganainy2010NP,Yang2014NP,Brandstetter2014NC,Khajavikhan2014S,Schindler2011PRA,Lin2011PRL,Liertzer2012PRL,Zhong2018PRA,Rotter2013PRX,Castaldi2013PRL,Fleury2018NP,Yang2018NPhon,Feng2013NM,Feng2014S,Longhi2009PRL,Chen2018NE}. This however comes at a price: isolated EPs are very sensitive to unavoidable fabrication errors or experimental uncertainty (e.g. small variation in the experimental conditions). To better appreciate this point, consider the current implementations of photonic EPs based on PT-symmetric coupled elements \cite{El-Ganainy2010NP,Yang2014NP,Brandstetter2014NC,Khajavikhan2014S} or engineered back reflection \cite{Wiersig2011PRA,Wiersig2014PRL,Wiersig2016PRA,Yang2017N}. In both of these geometries, which have been recently exploited to demonstrate ultra-responsive optical sensors \cite{Yang2017N,Khajavikhan2017N}, the design parameters have to be tailored precisely in order to force the system to operate at an EP. In the PT-symmetric implementation \cite{Khajavikhan2017N}, the resonant frequencies of the two rings have to be identical; the gain/loss profiles have to be exactly balanced; and the difference between the gain and loss values has to match the coupling coefficient between the two resonators. Alternatively, in the single ring implementation \cite{Wiersig2011PRA,Wiersig2014PRL,Wiersig2016PRA,Yang2017N}, the sizes and locations of the nanoscatterers next to the ring have to be controlled with high precision during the fabrication. To overcome these difficulties, various research teams employ clever techniques (such as micro-heaters and movable fiber tips, tunable coupling, etc) in order to actively and continuously tune the studied systems in the vicinity of the EPs. Beyond these important proof-of-concept demonstrations, it will be extremely useful for practical sensing applications to advance new design concepts that decouple the effects of fabrication errors and experimental uncertainties from perturbations caused by measurements.    

\begin{figure}
	\includegraphics[width=3.4in]{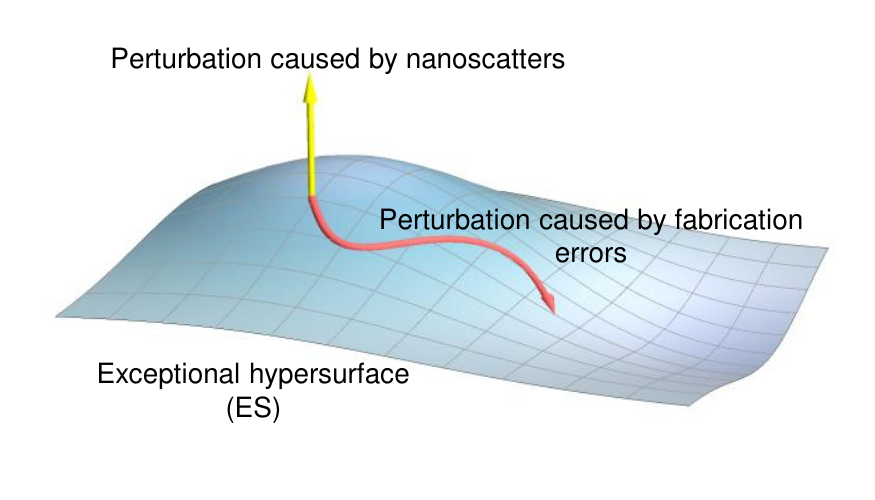}
	\caption{A non-Hermitian photonic structure can combine robustness together with sensitivity if it exhibits a hypersurface of exceptional points (ES) with the following properties: (1) Undesired perturbations due to fabrication imperfections and experimental uncertainties shift the spectrum across the surface, leaving the system at an EP; (2) Perturbations accounting for the quantities to be measured forces the spectrum out of the surface, i.e. away from EPs.}
	\label{FigES} 
\end{figure}

In this work, we present a new non-Hermitian photonic structure that exhibits an exceptional hypersurface (ES) embedded in a higher dimension parameter space. This, in turn, provides additional degrees of freedom that can be exploited to combine robustness with enhanced sensitivity. Particularly, robustness can be achieved if the system's response is tailored such that a large class of fabrication errors and experimental uncertainties shift the operating point along the ES. On the other hand, enhanced sensitivity can arise if the perpetration due to the measurements forces the spectrum away from the ES, causing large splitting of the resonant frequency (as compared to that associated with diabolic points \cite{Kippenberg2002OL,Kippenberg2009PRL,Yang2010PRA,Ozdemir2012OE,Yang2010NP,Ozdemir2014PNAS,Vollmer2015AOP}). This generic concept is illustrated schematically in Fig. \ref{FigES}. Here we show that this concept can be implemented by using standard photonic technology, which paves the way towards practical applications of non-Hermitian photonic sensors. 

\section*{Results}

\begin{figure*}
	\includegraphics[width=3.4in]{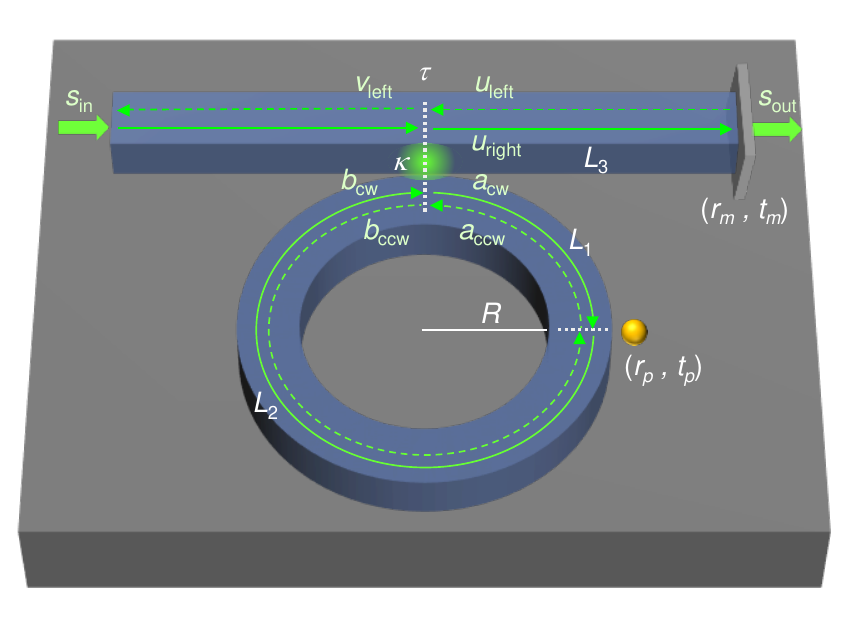}
	\caption{Schematic diagram of the proposed photonic structure that satisfies the criteria mentioned in Fig. \ref{FigES}. It consists of a microring resonator coupled to a waveguide that has a mirror on one side and reflectionless at the other end. The relevant design parameters are indicated in the figure (see Supplementary Note A for details). In the absence of any reflective perturbations, the system exhibits an EP. Any variations of the coupling coefficients or the resonant frequency of the cavity will still leave the system at an EP. On the other hand, if a nanoscatterer (or any other form of reflective perturbations) come to the vicinity of the ring, it will introduce a bidirectional coupling between the clockwise (CW) and counterclockwise (CCW)  waves and shifts the system away from the EP which in turn will leave a fingerprint on the emission spectrum of the system (if used in the lasing regime) or the power scattering spectrum (if operated in the amplification regime).}
	\label{FigSchematic} 
\end{figure*}

To this end, we consider the structure depicted schematically in Fig. \ref{FigSchematic}. It consists of a single microring resonator coupled to a waveguide. One end of this waveguide is terminated by a mirror while the other end is assumed to be reflectionless (we will discuss the effect of finite small reflectivity later, see Appendix A). Within the context of coupled mode theory (CMT), the above structure in the absence of the scatterer can be described by the effective Hamiltonian: 
\begin{equation} \label{H_ES}
i\frac{d}{dt} \begin{bmatrix}
\tilde{a}_\text{cw} \\ \tilde{a}_\text{ccw}
\end{bmatrix}
=H_\text{ES} \begin{bmatrix}
\tilde{a}_\text{cw} \\ \tilde{a}_\text{ccw}
\end{bmatrix},
H_\text{ES}=
\begin{bmatrix}
\omega_0 -i \gamma   & 0  \\
 \alpha \mu^2    & \omega_0 -i \gamma 
\end{bmatrix}
\end{equation}
where $\tilde{a}_\text{cw,ccw}$ are the field amplitudes of the clockwise (CW) and counterclockwise (CCW) modes, $\omega_0$ is the resonant frequency, $\gamma$ is the cavity loss rate which can be decomposed into intrinsic absorption, radiation loss, and loss to the waveguide (i.e. $\gamma=\gamma_\text{abs}+\gamma_\text{rad}+\mu^2/2$), and $\mu$ is the coupling rate between the resonator and the waveguide. In addition, $\alpha=r_m \exp(i2\phi_3)$ where $r_m$ is the field reflection coefficients at the mirror and $\phi_3=\beta_w L_3$. Here $\beta_w$ is the propagation constant of the waveguide and the distances $L_3$ are depicted in Fig. \ref{FigSchematic}. Note that the above form of the Hamiltonian does not necessarily imply that the system is nonreciprocity.

The eigenvalues of $H_\text{ES}$ as written in the bases $\exp(-i\omega t)$, together with the associated eigenvectors $\tilde{\mathbf{a}}_{1,2}$ are given by:
\begin{equation}
\begin{gathered}
\begin{split}
&\omega_{1,2}=\omega_0 - i\gamma, 
\\
&\tilde{\mathbf{a}}_{1,2}=(0 , 1)^T.
\end{split}
\end{gathered}
\end{equation}

The spectrum of the Hamiltonian $H_\text{ES}$ features an EP with two identical eigenmodes characterized by a finite CCW component and a null CW component. Importantly, this is even true for any value of $\omega_0$, $\gamma$ and $\alpha \mu^2$. In other words, there is hypersurface spanned by all possible values of these parameters where the system remains at an EP. For instance, if the fabricated system has extra (less) loss, stronger (weaker) coupling to the waveguide or a shift in its resonance frequency from the original targeted values, the system will be still located at an EP without the need for any external tuning. Only perturbations that introduce differential loss or frequency mismatch between the two modes (CW and CCW) can affect the system performance. However, these perturbations do not arise naturally in our proposed design since any change in the shape/size of the resonator or its coupling to the waveguide will affect both modes symmetrically. This unique feature provides unprecedented robustness that cannot be achieved in standard non-Hermitian systems that rely on isolated EPs in the design parameter space. However, in the presence of a nanoscatterer located in the vicinity of the ring resonator, the interaction between the scatterer and the evanescent field of the optical modes introduces a bidirectional coupling between the CW and CCW modes, which are described by additional corrections of the same order to both off-diagonal matrix elements of $H_\text{ES}$, say $\epsilon$. If we further assume that $\epsilon$ is much smaller than other matrix elements, it is straightforward to show that the splitting of the eigenfrequency is $\Delta \omega \equiv |\omega_1-\omega_2| \sim \sqrt{\epsilon}$. In standard waveguide-coupled microring resonators operating at a DP, this splitting will be rather $\epsilon$. Thus, in addition to its robustness, the proposed system is expected to also provide enhanced sensitivity. 

In order to put this discussion on a more solid ground while at the same time elucidate on the relevant experimental parameters, we study the above structure using the scattering matrix method (SMM) \cite{Yariv2000EL,Van-OMR,Saleh-FP}. Here we assume that the system is probed via the waveguide by a signal $s_\text{in}$. We then proceed to calculate the output signal $s_\text{out}$ as a function of the input frequency for different levels of perturbations by a nanoscatterer, which we quantify by its location as well as reflection/transmission coefficients $r_p$/$t_p$, respectively (see Fig. \ref{FigSchematic} and  Appendix B for a full list of parameters).  

By doing so, we obtain (see Appendix B for details):
\begin{equation} \label{eq:sout}
\frac{s_\text{out}}{s_\text{in}}=\dfrac{e^{i \phi_3} t_m [(1+e^{2i \phi}) \tau-e^{i \phi}(1+\tau^2)t_p]}
{1+e^{i 2 \phi} \tau^2-2 e^{i \phi}\tau t_p - e^{2i\phi'} r_m r_p \kappa^2}
\equiv \dfrac{N}{D},
\end{equation}
where $\phi=\phi_1+\phi_2$ and $\phi'=\phi_2+\phi_3$, with $\phi_{1,2}=\beta_r L_{1,2}$ and $\phi_{3}=\beta_w L_{3}$. In general, the values of propagation constants associated with the ring and straight waveguides, $\beta_{r,w}$, can be complex with the imaginary parts accounting for the possible radiation and material loss as well as the gain (loss due to coupling to the waveguide is treated separately). For reasons that will be clear shortly, we are particularly interested in the case of active devices where the microring exhibits enough optical gain to bring the system at or close to the lasing condition (for completeness we treat the passive case in Appendix C). Under either of these conditions,  the lasing or the transmission spectrum (respectively) is dominated by the the poles of the power scattering coefficient $T=|s_\text{out}/s_\text{in}|^2$, or equivalently the zeros of $D$. To characterize the performance of the proposed structure, we thus study the behavior of $D$ as a function of the particle reflectivity $r_p$ and the input frequency parametrized by $\phi$ (we do not take the waveguide dispersion into account at this moment), i.e. $D\equiv D(r_p,\phi)$. 

For any set of design parameters and a specific value of $r_p$, the lasing conditions is achieved for values of $\phi \equiv \phi_D$ satisfying the equation $D(r_p,\phi_D)=0$, which gives $\exp(i \phi_D^{\pm})=\tau^{-1} (t_p \pm i\sqrt{r_p^2-\exp(2i \phi') r_m \kappa^2 r_p })$. The maximum frequency splitting $\Delta \phi \equiv \text{Re}[\phi_D^{+}-\phi_D^{-}]$ occurs when  $\exp[2i \text{Re}(\phi')]=-1$. As a side comment, we note that $\text{Im}[\phi_D]=-\kappa^2/2$, which implies the lasing threshold occurs when the gain is enough to compensate for the radiation/material loss as well as the loss due to coupling to the waveguide, as one would expect. By writing $r'_m=r_m \times |\exp(2i\phi')|$ we find: 
\begin{equation} \label{Eq:Splitting}
\Delta \phi =2\sqrt{r_p^2+r'_m \kappa^2 r_p} 
\end{equation}
In Eq. (\ref{Eq:Splitting}), $r'_m\kappa^2$ is the effective unidirectional coupling from CW mode to CCW mode. By noting that in our systems, both $r_m$ and $|\exp(2i\phi')|=\exp(-2 \text{Im}[\phi'])$ are in the order of unity (since the system is assumed to operate below but close to the lasing threshold), we arrive at:
\begin{align} \label{Eq:EPSplitting}
\Delta \phi_\text{EP}
&\approx \left\{\begin{matrix}
&2\kappa \sqrt{r_p}, & r_p \ll \kappa^2 \\
&2r_p, & r_p \gg \kappa^2
\end{matrix} \right. ,\\
\label{Eq:DPSplitting}
\Delta \phi_\text{DP} &=  2r_p .  
\end{align}

Equation (\ref{Eq:EPSplitting}) is the central result of this work. It confirms the existence of an operating regime ($r_p \ll \kappa^2$) where the frequency splitting scales with the square root function of the perturbation, which is the hallmark of enhanced sensitivity near a second order EP. Beyond this regime, the splitting is linear as in standard sensors operating at a diabolic point. Intuitively, as the perturbation due to the scatterer shifts the system far away from the EP, the extra sensitivity is lost. 

In the active scattering regime, when the gain brings the system relatively close to the lasing point but remains below the lasing threshold, the transmission peaks can be obtained by solving Eq. (\ref{eq:sout}). Not surprisingly, here also the locations of the transmission peaks are dominated by the zeros of $D(r_p,\phi)$, which again results in a square-root dependence of the frequency splitting as we have confirmed numerically.   

\begin{figure*} [!b]
	\includegraphics[width=4.5in]{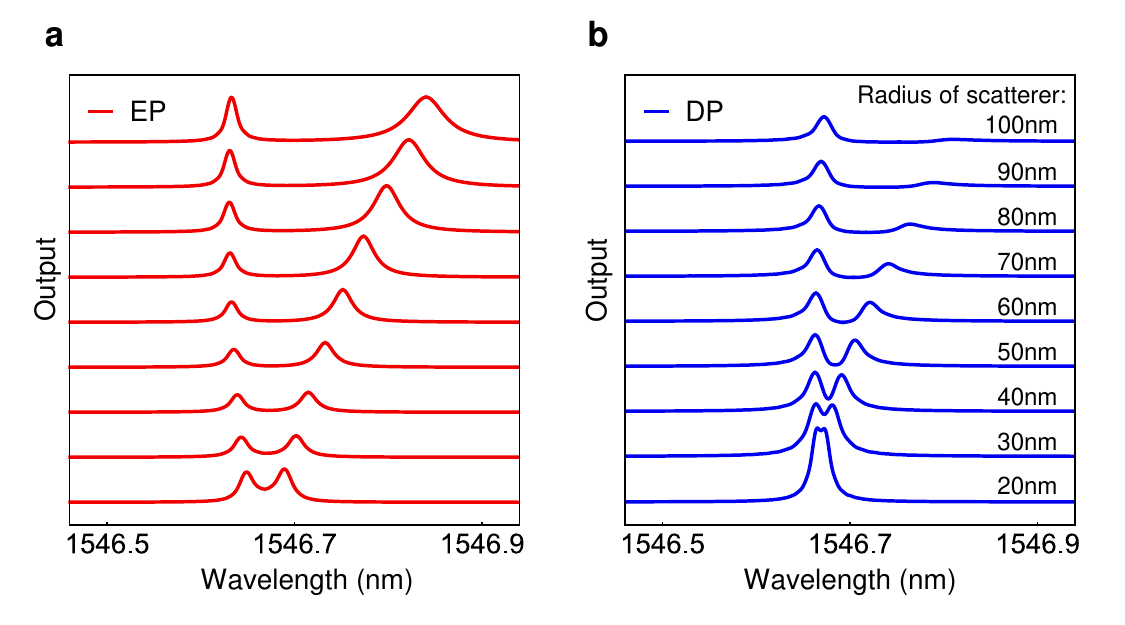}
	\caption{ Finite difference time domain simulations for a system similar to that of Fig. \ref{FigSchematic}. The width of the straight and ring waveguides is $w$ = 0.8 $\mu$m and the ring radius is $R$ = 10 $\mu$m. The separation between the waveguide and ring is $d$ = 0.6 $\mu$m. To simulate the perturbation induced by a nanoscatterer, we use a disk and vary its radius in the simulation from 20 nm to 100 nm. The disk is located at 3 o'clock with a fixed distance $h$ = 0.1 $\mu$m from its center to the outside of the microring. A 50-nm-thick and 4-$\mu$m-wide Ag is used as a mirror on the right side of the waveguide with a distance of $L_3$ = 10.075 $\mu$m. Note that the location of the nanoscatterer and the mirror are chosen to optimize the performance. Finally, The waveguide, microring resonator, and the scatterer have same real part of the refractive index $n_2$ = 1.45, while the microring resonator exhibits an additional imaginary part of $-4.7\times10^{-5}$ to account for the gain. The system is excited by a broad bandwidth pulse launched at the left side of the waveguide, and the transmission spectrum is measured at the right port.  \textbf{a} and \textbf{b} plot the spectrum splitting as a function of nano-scatterer size. Clearly, the EP-based structure demonstrates superior performance in terms of the splitting magnitude and the visibility of the resonance peaks.}
	\label{FigFDTD} 
\end{figure*}

Having discussed the essential features of the proposed structure, we now confirm our predictions by performing two-dimensional full-wave simulations \cite{Taflove-CE} using realistic material platforms. Figure \ref{FigFDTD}a depicts the simulated geometry. It consists of a microring resonator having a refractive index $n_2$ = 1.45, a radius $R$ = 10 $\mu$m, and a width $w$ = 0.8 $\mu$m. The ring is coupled to a waveguide having the same material and width. The edge-to-edge separation between the ring and the waveguide is chosen to be $d$ = 0.6 $\mu$m, corresponding to $\kappa^2$ = 0.028. A mirror with reflectivity $r_m$ = 0.99 is introduced at one end of the waveguide via a 50-nm-thick silver layer. The nanoscatterer is assumed to be of the same material. Based on the chosen position of the nanoscatterer (see Fig. \ref{FigSchematic}), then we chose $L_3$=10.075$\mu$m in order to achieve optimal operation 
(defined with respect to the splitting associated with a test nanoscatter of radius 30 nm). Finally, the background material is assumed to be air of $n_1=1$. In our simulations, the device is probed by a TE-polarized broad bandwidth pulse with central frequency at $f$ = 193.4 THz or equivalently $\lambda$ = 1550 nm (almost matching one of the longitudinal modes of the microring) launched from the left side of the waveguide. In order to isolate the relevant transmission peaks in our simulations, we used a dispersive gain function as described in Appendix D.

\begin{figure*} [!b]
	\includegraphics[width=5.8in]{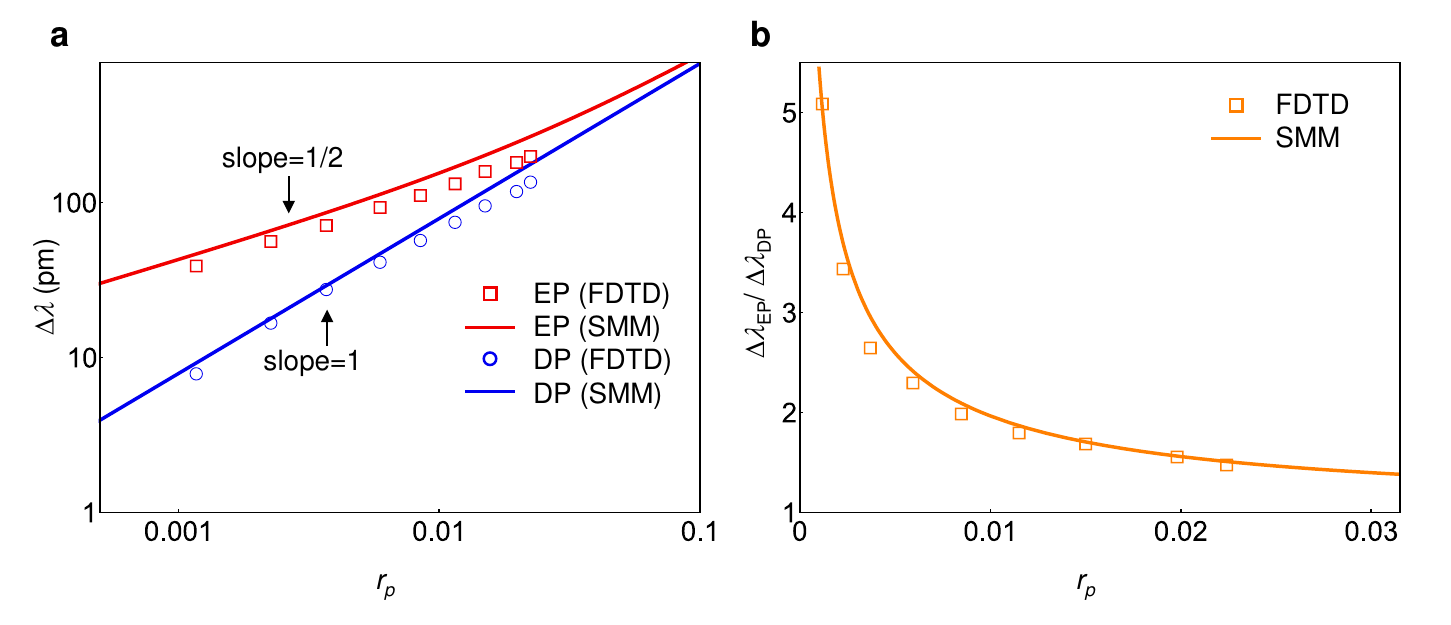}
	\caption{Sensitivity enhancement as a function of the nanoscatterer radius. Clearly the EP sensor has a better performance than a sensor operating at a DP for smaller scatterer, making this device valuable for measuring small perturbations.}
	\label{FigEnhancement} 
\end{figure*}

Figures \ref{FigFDTD}a and \ref{FigFDTD}b show the transmission spectrum for the cases of EP and DP, respectively for the parameters listed in the figure caption. Evidently, the EP-based device exhibits a significant advantage, demonstrating larger splitting and clear transmission peaks. Note that the location of one transmission peak remains almost invariant while the other experience red-shift. This can be explained by noting that the field profile corresponding to these peaks is that of a standing wave with the particle positioned at the dark and bright fringes in both cases respectively. Figure \ref{FigEnhancement}a plots a log-log scale of the slopes characterizing the magnitude of the splitting, where the superior performance of EP is evident. This conclusion is better illustrated in figure \ref{FigEnhancement}b, which depicts the enhancement factor (defined as the ratio of the splitting in the EP case normalized by that of the DP case) of the proposed sensor as a function of the nanoscatterer reflectivity $r_p$ when the scatterer size is varied from 20 nm to 100 nm (see Appendix E).  These figures also demonstrate the excellent agreement between the FDTD (square points) and  the scattering matrix method (solid lines).  

\section*{Discussion}
In conclusion, we have proposed a new class of non-Hermitian sensors that operate at exceptional surfaces as opposed to isolated exceptional points. This new paradigm provides more degrees of freedom that can be exploited to combine the robustness needed for real-life applications together with the enhanced sensitivity associated with exceptional points. Towards this goal, we have also proposed a specific device design that can be implemented by standard photonics technology. To establish the superior performance of the proposed structure, we have investigated its operation by using the scattering matrix formalism, showing that it can function both in the lasing and amplifying modes. These predictions have been confirmed by performing two-dimensional full wave simulations using a realistic material platform. Additionally, we also expect our proposed system to demonstrate some robustness against the type of thermal fluctuations studied recently in \cite{Wolff2018arXiv}. In fact, any temperature variation will shift the central frequency of both CW/CCW modes isotropically but will not affect the splitting (as has been previously shown for the case of DPs \cite{Yang2010APL}. As we discuss in Appendix F, initial estimations indicate that the operation will not affect if the temperature vary by $\sim \pm 20 {\text{K}}$. Another important aspect is the performance of this device in the quantum limit. Very recently, it was shown that non-Hermitian Hamiltonians with unidirectional coupling can exhibit superior performance in the quantum regime \cite{Lau2018NC}. We anticipate that our results will open a host of new possibilities for sensing applications using practical non-Hermitian devices. Importantly, the proposed design concept presented here can be implemented in other physical platforms such as acoustics or microwaves.

\section*{Appendix}
\appendix
\renewcommand\thefigure{A\arabic{figure}}
\setcounter{figure}{0}  

\section{Eliminating reflection from input port}
As we discussed in the main text, an important condition for our proposed system to function properly is to eliminate or minimize (compared to other parameters) the reflection from the input port. In the context of laser engineering, various techniques have been developed to minimize port reflection. These include cleaving the waveguide end at slanted angles \cite{Mundbrod2003,Chen2017OIT} and using anti-reflection coating \cite{Kaminow1983JQE}. Power reflection values as low as $\sim 10^{-7}$ have been reported \cite{Mundbrod2003}. In addition to these methods, one can also utilize absorption. For instance, if the far end of the waveguide is coupled to a resonator that has a high loss, then, the total reflection can be further reduced by orders of magnitude. The use of optical circulators with high isolation ratios is another approach. Recently, a novel technique based on Kramers-Kronig relation was also proposed to eliminate reflections  \cite{Ye2017NC}. We plan to explore these different design strategies in future work.

\section{Scattering matrix method}
Here we present more details on the derivation of Eq. (3) in the main text. The scattering matrices associated with the evanescent coupling region $S_c$, the partially reflective mirror, and nanoscatterer are given by:
\begin{equation}  \label{Eq:Scattering matrix}
S_c=\begin{bmatrix}
\tau & i\kappa \\
i\kappa & \tau
\end{bmatrix},
~S_m=\begin{bmatrix}
t_m & i r_m \\
i r_m & t_m
\end{bmatrix},
~S_p=\begin{bmatrix}
t_p & i r_p \\
i r_p & t_p
\end{bmatrix},
\end{equation}
where $\kappa$ is the coupling between the waveguide and ring resonator, $\tau$ is the transmission and they satisfy $\kappa^2+\tau^2=1$ (assuming no loss); $r_{m,p}$  and  $t_{m,p}$ are the reflection and transmission coefficients of the mirror and scatterer, respectively, and satisfy $r_{m,p}^2+t_{m,p}^2=1$ (again assuming no loss). All these parameters are real positive numbers.  

The electric fields components of Fig. 2 in the main text, with the internal components calculated at the coupling regions (dashed lines of Fig. 2) are then given by \cite{Van-OMR}:
\begin{equation}
\begin{aligned}
\begin{bmatrix}u_\text{right} \\ a_\text{cw}\end{bmatrix}
&=S_c \begin{bmatrix}s_\text{in} \\ b_\text{cw}\end{bmatrix}, \\
\begin{bmatrix}v_\text{left} \\ b_\text{ccw}\end{bmatrix}
&=S_c \begin{bmatrix}u_\text{left} \\ a_\text{ccw}\end{bmatrix},  \\
\begin{bmatrix} b_\text{cw} \exp(-i\phi_2) \\ a_\text{ccw} \exp(-i\phi_1)\end{bmatrix} 
 &=S_p \begin{bmatrix} a_\text{cw} \exp(i\phi_1)\\ b_\text{ccw} \exp(i\phi_2)\end{bmatrix},  \\
\begin{bmatrix} s_\text{out}\\ u_\text{left} \exp(-i\phi_3)\end{bmatrix}  
&=S_m  \begin{bmatrix} u_\text{right} \exp(i\phi_3) \\ 0 \end{bmatrix}. \\
\end{aligned}
\end{equation}  
Note that material and radiation loss as well as gain can be incorporated in the imaginary components of the phases $\phi_{1,2,3}$. It is straightforward to drive the formula for the transmission spectrum from the above set of equations.

\section{Passive EP sensors}
In the main text, we considered the operation of the system only in the amplification and lasing regimes. Here, for completeness, we also analyze the passive structure, i.e. when no gain is applied. Under this condition, the transmission spectrum is dominated by the zeros of numerator term $N$ (see Eq. (3) of the main text), denoted by $\phi_N$, which correspond to dips in the transmission and given by: 
\begin{equation}
\exp(i \phi_N)=K t_p \pm \sqrt{K^2 t_p^2-1},
\end{equation}
where $K=(\tau+1/\tau)/2$. By recalling that  $\kappa=\sqrt{1-\tau^2}$, and noting that in our system $\kappa \ll 1$ (in fact, $\kappa \sim 0.167$ in our FDTD simulations), we find that  $K \sim 1+\frac{\kappa^4}{8}+O(\kappa^6)$, which leads to a complex frequency  splitting $\phi_N^{\pm}=\pm \sqrt{r_p^2-\kappa^4/4}$. Thus, only when $r_p > \kappa^2/2$ , $\Delta \phi_N \equiv \phi_N^{+}- \phi_N^{-}=2\sqrt{r_p^2-\kappa^4/4}$  is real and correspond to two dips in the transmission spectrum. When $r_p > \kappa^2/2$, the complex splitting correspond to a change in the modal lifetime \cite{Kippenberg2010NP}. These results applies equally for DPs and EPs.

The situation becomes drastically different if we consider the reflected signal instead of the transmitted one as we have done so far. In this case, we are interested in the quantity: 
\begin{equation}  
\frac{v_\text{left}}{s_\text{in}}=\dfrac{i e^{-2i\phi_3} [e^{2i\phi}r_m-2e^{i \phi}t\tau r_m +\tau^2 r_\text{m} -e^{2i(\phi-\phi')}r \kappa^2]}
{1+e^{i 2 \phi} \tau^2-2 e^{i \phi}\tau t_p - e^{2i\phi'}r_m r_p \kappa^2}.
\end{equation}
By assuming $r_m \sim1$ and $\exp(2i\phi') \sim-1$, the reflection vanishes (i.e. $v_\text{left}=0$) when:
\begin{equation}
\exp(i\phi_N)=\frac{\tau}{1+r_p \kappa^2}(t_p \pm i \sqrt{r_p^2+\kappa^2 r_p})
\end{equation}
As expected, $\text{Im}(\phi_N)=\frac{\kappa^2}{2}$ which gives the condition for critical coupling \cite{Yariv2002PTL}. Surprisingly however, in this case, we have $\text{Re}(\phi_N)=\pm \sqrt{r_p^2+\kappa^2 r_p}$. Thus for very small values of $r_p$, the splitting is $\sim 2\kappa \sqrt{r_p}$ as in the active device. 

\section{Gain dispersion}
The material gain in the FDTD simulation is assumed to depend on the light frequency via  a Lorentz dispersion model: 
\begin{equation}
\varepsilon_\text{total}(\omega)=\varepsilon+\frac{\varepsilon'\omega_0^2}{\omega_0^2-2 i \delta \omega-\omega^2}
\end{equation}
with $\varepsilon=1.45^2$; $\omega_0=1.2179 \times 10^{15}$ rad/s (corresponding to $\lambda$ = 1546.7 nm); $\delta=2\times 10^{12}$ rad/s; $\varepsilon'=-4.5\times10^{-7}$ corresponds to an imaginary index around $-4.7\times 10^{-5}$ at resonance. 

\section{Reflection from nanoscatterer} 
\begin{figure}[!htb]
	\includegraphics[width=3in]{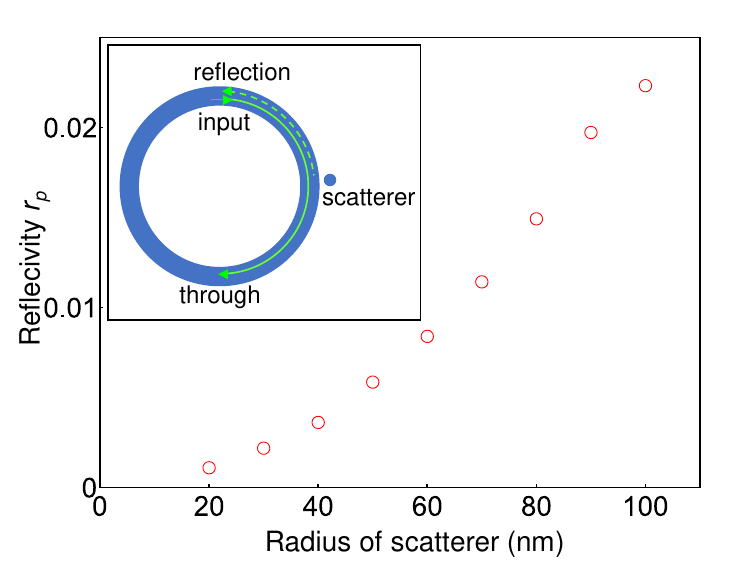}
	\caption{An electric field with wavelength 1550 nm is injected from the top of the ring resonator. The reflection component is measured on the left side of the input and the reflectivity is obtained.    }
	\label{FigParticleReflection} 
\end{figure}

In our scattering matrix formalism, the scatterer was quantified via its reflection $r_p$ and transmission $t_p$ coefficients, respectively. However in the FDTD simulations, we varied the scatterer size. In order to compare the two methods, we need to establish the relation between these two parameters. To do so, we used the geometry shown in the inset of Fig. \ref{FigParticleReflection}. The values of $r_p$ as a function of the scatterer radius are also presented.

\section{Temperature sensitivity} 
In addition to  being robust against fabrication errors, our proposed device can also withstand wide range of temperature variation, which can be quickly checked by noting that the thermal coefficient of silica $dn/dT$ is around $12\times10^{-6}$ K$^{-1}$ \cite{Toyoda1983JPD}.  Assuming a ring resonator of radius $R$ = 10 $\mu$m and a temperature change is $\Delta T$ = 20 K. The estimated length $L_2+L_3$ corresponding to $\phi'$ can be equal to the perimeter of the ring,  the shift in the value of $\phi'$ is:
\begin{equation}
\Delta\phi'=\dfrac{4 \pi^2 R \frac{dn}{dT}\Delta T}{\lambda}=0.06
\end{equation}
which is very small and does not affect the performance significantly.  Note that the thermal expansion of silica, $\approx 0.5\times10^{-6}$ K$^{-1}$, can be safely neglected.

Thus our analysis shows that a temperature variation within $\pm$20 K around the optimal operation point can be easily tolerated by our device. \\


\end{document}